\documentstyle[11pt,epsfig]{article}  
\begin{document} 

%%%%%%%%%%%%%%%%%%%%%%%% NEW DEFINITIONS 
\def\la{\mathrel{\mathpalette\fun <}}
\def\ga{\mathrel{\mathpalette\fun >}}
\def\fun#1#2{\lower3.6pt\vbox{\baselineskip0pt\lineskip.9pt
\ialign{$\mathsurround=0pt#1\hfil##\hfil$\crcr#2\crcr\sim\crcr}}}  
\def\lrang#1{\left\langle#1\right\rangle}
%%%%%%%%%%%%%%%%%%%%%%%% END OF NEW DEFINITIONS 

%%%%%%%%%%%%%%%%%%%%%%%% TITLE PAGE

\begin{titlepage} 
\begin{center}
{\Large \bf Photon-jet correlation in heavy ion collisions at the LHC~\footnote{ 
Contribution to the CERN Yellow Report on Hard Probes in Heavy Ion 
Collisions at the LHC.}} 

\vspace{5mm}

{\it O.L.~Kodolova$^a$, I.P.~Lokhtin$^a$ and A.~Nikitenko$^b$} 

\vspace{2.5mm} 
    
$^a$ Institute of Nuclear Physics, Moscow State University, Russia \\ 
$^b$ Imperial Colledge, London, UK   

\vspace{5mm} 

\end{center}  

\begin{abstract} 

Transverse momentum imbalance between a jet produced with a hard photon is 
considered as a signal of medium-induced partonic energy loss in 
ultrarelativistic heavy ion collisions. We analyze photon-jet correlation in 
the context of a real experimental situation at the LHC, the problem of 
neutral pion background being discussed.  

\end{abstract}

\end{titlepage}
 
\begin{figure}[hbtp]
\begin{center} 
\resizebox{90mm}{90mm} 
{\includegraphics{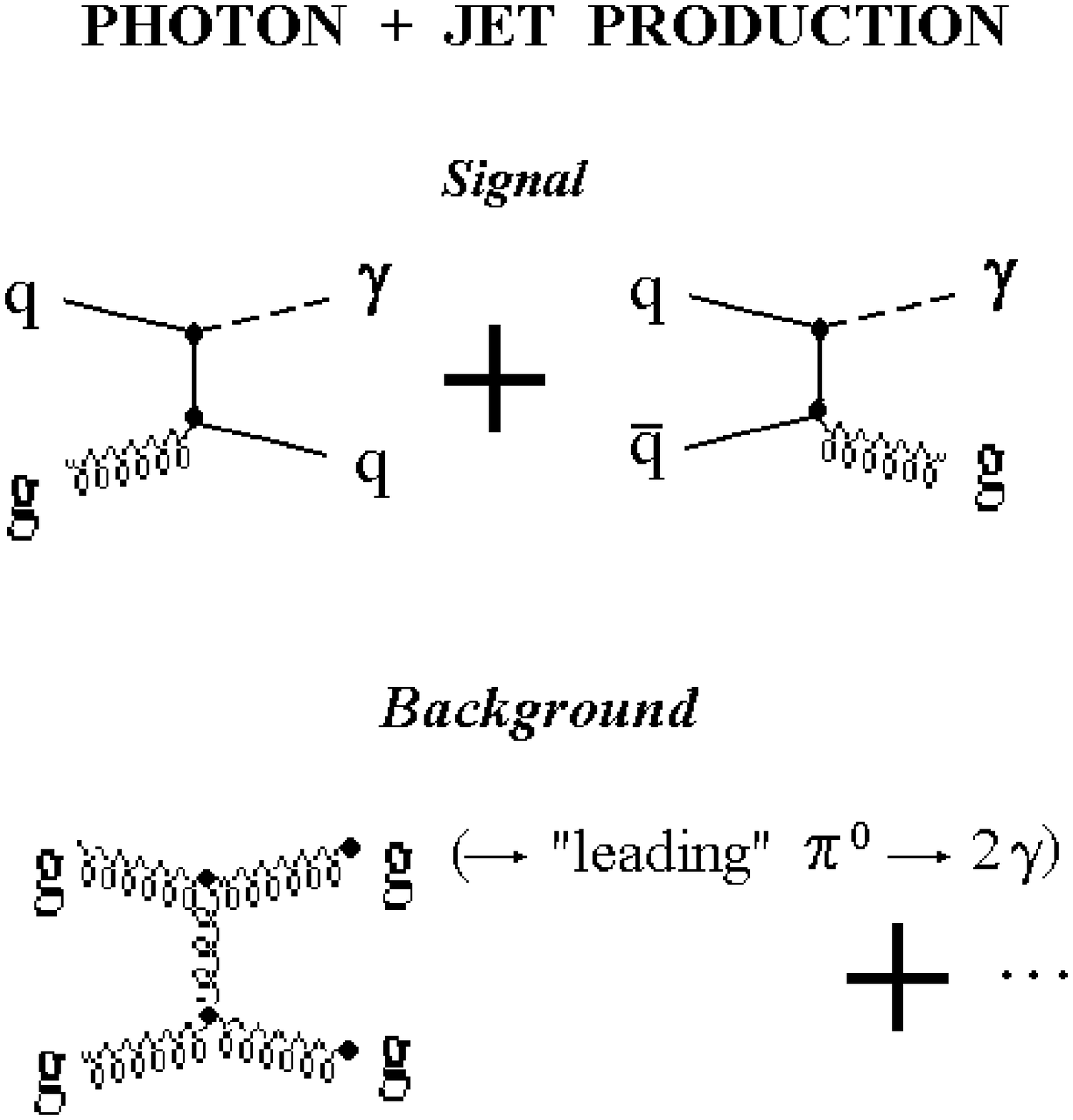}} 
\caption{\small Leading order diagrams for  $\gamma$+jet (signal) and  $\pi_0 (\rightarrow 
2\gamma)$+jet (background) production. }
\label{gamjet:fig1}

\vskip 1 cm 

\resizebox{80mm}{100mm} 
{\includegraphics{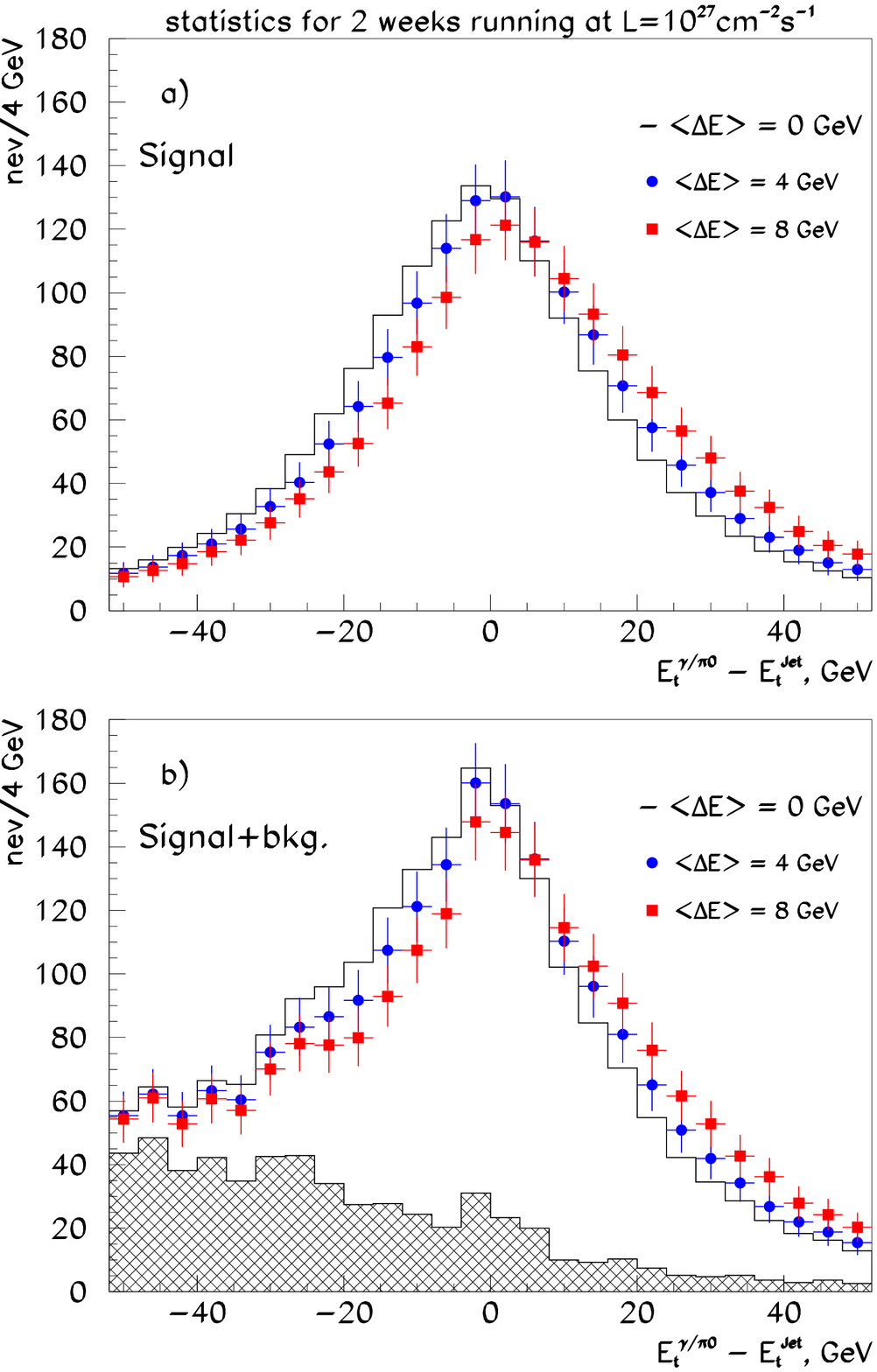}} 
\caption{\small The distributions of differences in transverse energy between the $\gamma$ and 
jet with $E_t^{\gamma,~jet} > 120$ GeV: a) without ($\pi^0$ + jet) background,  b) with 
($\pi^0$ + jet) background. The pseudorapidity coverage is  $\mid \eta_{\gamma,~jet} \mid < 1.5$. 
Different values of jet in-medium energy loss, initial state gluon radiation and finite jet energy
resolution are taken into account.}
\label{gamjet:fig2}
\end{center} 
\end{figure}

Among with other proposed signals of jet quenching, the 
$P_T$-imbalance between a produced jet with a gauge boson in $\gamma+$jet~\cite{Wang:1996}
and $Z+$jet~\cite{Kartvelishvili:1996} production has been identified as being observable in 
heavy ion collisions at the LHC~\cite{Baur:2000}.  The dominant leading order 
diagrams for high transverse momentum $\gamma$ + jet production are shown in 
figure~\ref{gamjet:fig1}. Contrary to the gluon-dominated jet pair production where one could 
investigate jet quenching due to mostly gluon energy loss in dense matter,  $\gamma+$jet 
channel gives a possibility to study quark energy loss.  The main background here is hard jet pair 
production when one of the jet in an event is misidentified as a photon. The leading $\pi ^{0}$ in 
the jet is a main source of the misidentification. Table 1 presents the event rates 
for signal and background processes in one month of Pb$-$Pb beams (a half of the time is 
supposed to be devoted to data taking), $R= 1.2 \times 10^6$ s, assuming luminosity 
$L = 5 \times 10^{26}~$cm$^{-2}$s$^{-1}$ to that $$N(events)= R \sigma^{h}_{AA} L ,$$ where 
production cross section in minimum bias nucleus-nucleus collisions were obtained from those in 
$pp$ interactions at the same energy ($\sqrt{s} = 5.5$ TeV) using simple parameterization 
$\sigma^h_{AA}=A^2 \sigma^h_{pp}$. The pseudorapidity acceptance of CMS
experiment is considered. The cross section in $pp$ collisions were evaluated
using the PYTHIA\_$6.1$ Monte-Carlo generator~\cite{pythia} with the CTEQ5L parton 
distribution function. Note that the influence of nuclear shadowing is practically negligible for the 
region of  sufficiently hard $\gamma+$jet production, $x_{1,2} \sim \sqrt{\widehat{s}/s} \ga 
0.2$. However, large theoretical uncertainties in absolute rates in $pp$ collisions come 
from choice of the parton distribution functions, next-to-leading corrections, etc. 
It means that measurements in $pp$ or $dd$ collisions at 
the same or similar energies per nucleon as in the heavy ion runs are strongly desirable to 
determine the baseline rate precisely. 

\begin{table}[htb]
\begin{center}
\label{gamjet:tab1} 
\caption{\small Expected rates for $\gamma$+jet and  $\pi_0 (\rightarrow 2\gamma)$+jet channels 
in one month of Pb$-$Pb beams.} 

\medskip 

\begin{tabular}{|l|c|c|} \hline  
Channel & $|\eta|<1.5$ & $|\eta|<3$ \\ \hline 
$\gamma$+jet, $E_T^{jet,\gamma}>100$ GeV & 1.6$\times$10$^3$ & 3.0$\times$10$^3$  
\\ \hline 
$\pi_0 (\rightarrow 2\gamma)$+jet, $E_T^{jet,\gamma}>100$ GeV & 8.4$\times $10$^3$ 
& 2.2$\times$10$^4$  \\ \hline  
\end{tabular}
\end{center}
\end{table}

One can see that even for events with $E_T>100$ GeV the
background is still dominant. Signal-to-background ratio becames close to $1$ only above
$200$ GeV. The identification of the influence of the dense medium formation on signal spectra
requires the reduction of the background. One of the possibilities is to apply some kind of the
photon isolation, so called "zero suppression criteria", which requiers no energy above a given
threshold around photon (see section "Photon detection at CMS" for details).  In this case
signal-to-background ratio at $E_T>100$ GeV can be improved by factor about 2.3 by the price of
$14\%$ of the signal reduction. Other possibility is to apply some kinematical cuts, which do not
have influence on the $P_T$-imbalance of the process (i.e. not result in shift of maximal value of 
$E_T^{\gamma}-E_T^{jet}$ distribution).  

The possibility to observe the medium-induced energy loss of quark-initiated jet 
using photon-jet correlation in heavy ion collisions with CMS detector has been 
investigated in~\cite{Kodolova2:1998,Baur:2000}. It has been found that initial state gluon
radiation and finite jet energy resolution (which is much larger than photon energy resolution) 
result in significant smearing of distribution of differences in transverse energies between the 
photon and jet. But it is still symmetric: $E_T^{\gamma}=E_T^{jet}$ only in average 
(not for each given event). The non-symmetric shape of the distribution appears if 
a jet loses energy: the maximal value of the distribution is equal to the average 
energy loss of the quark-initiated jet at given energy detection threshold, $E_T^{jet} \sim 100$
GeV in CMS case. Note  that we are not measuring energy loss of a leading quark by such a  
way, but getting total loss of quark-initiated jet outside the given jet cone. 

In order to test the sensitivity of $\gamma$+jet production to jet quenching, it was  
considered three scenarios with average collisional energy loss of a jet: 
$<\Delta E_{q}> \simeq 0$, $4$ and $8$ GeV respectively ($<\Delta E_{g}> = 9/4 \cdot  
<\Delta E_{q}>$). The jet energy resolution at mid-rapidity obtained for $Pb-Pb$ 
collisions has also been used to smear the energy of the recoiling parton, as well as 
the jet rejection factor and signal efficiency~\cite{Kodolova2:1998}. Figure~\ref{gamjet:fig2} 
shows the distributions of differences in transverse energy between the photon and 
jet with $E_T^{\gamma,~jet} > 120$ GeV  in one month of Pb$-$Pb beams without and  
with ($\pi^0$ + jet) background in the pseudorapidity region $\mid \eta^{\gamma,~jet} \mid < 1.5$ 
for different values of jet energy loss. In this case luminosity $L =  
10^{27}~$cm$^{-2}$s$^{-1}$ was assumed and PYTHIA\_$5.7$ version with the default CTEQ2L pdf
choice was used. The jet energy resolution leads to a difference between the input values 
$<\Delta E_{q}>$ and the ones obtained from the spectra. The background of 
$\pi^0$-contamination results in non-zero negative values of the final distributions 
(figure ~\ref{gamjet:fig2}b) already in the case without jet energy loss. However one can see on 
that shape of the distribution is well distinguish for the scenarios considered. For the region of 
$(E_T^{\gamma}-E_T^{jet}) > 0$ there is a difference for almost every bin greater than 
$1$ standard deviation for the rather small jet energy loss $8$ GeV and even for the loss $4$ GeV. 
In the real experiment it would be possible to estimate the number of background events using 
the region without the signal ($E_T^{\gamma}-E_T^{jet}) < - 100$ GeV/c and background shape 
from Monte-Carlo simulation and (or) from $pp$ data. Significant difference in the shape of 
$E_T^{\gamma}-E_T^{jet}$ distribution can allow to optimize extraction of the signal from
the experimental spectra.

It is pleasure to thank P.~Aurenche, M.~Bedjidian, D.~Denegri, L.I.~Sarycheva and U.~Wiedemann for 
encouraging and interest to the work. Discussions with R.~Kvatadze, P.~Levai, A.M.~Snigirev and 
I.N.~Vardanian are gratefully acknowledged.

\end{document}